\documentclass{emulateapj}

\usepackage{graphicx}
\usepackage{epsfig}
\usepackage{times}
\usepackage{natbib}
\usepackage{url}
\usepackage{color}
\usepackage{amssymb,amsmath}

\shorttitle{Fan-spine topology formation}
\shortauthors{T\"or\"ok et al.}

\begin{document}

\title{Fan-spine topology formation through 
two-step reconnection driven by twisted flux emergence}

\author{T. T\"or\"ok, G. Aulanier and B. Schmieder}
\affil{LESIA, Observatoire de Paris, CNRS, UPMC, Universit\'e Paris Diderot, 
       5 place Jules Janssen, 92190 Meudon, France}
\author{K.~K. Reeves and L. Golub}
\affil{Harvard-Smithsonian Center for Astrophysics, 60 Garden Street, 
       Cambridge, MA 02138, USA}

\begin{abstract}
We address the formation of 3D nullpoint topologies in the solar corona by 
combining {\em Hinode}/XRT observations of a small dynamic limb event, which 
occurred beside a non-erupting prominence cavity, with a 3D zero-$\beta$ MHD 
simulation. To this end, we model the boundary-driven ``kinematic'' emergence 
of a compact, intense, and uniformly twisted flux tube into a potential field 
arcade that overlies a weakly twisted coronal flux rope. The expansion of the 
emerging flux in the corona gives rise to the formation of a nullpoint at the 
interface of the emerging and the pre-existing fields. We unveil a two-step 
reconnection process at the nullpoint that eventually yields the formation of 
a broad 3D fan-spine configuration above the emerging bipole. The first 
reconnection involves emerging fields and a set of large-scale arcade field 
lines. It results in the launch of a torsional MHD wave that propagates along 
the arcades, and in the formation of a sheared loop system on one side of the 
emerging flux. The second reconnection occurs between these newly formed loops 
and remote arcade fields, and yields the formation of a second loop system on 
the opposite side of the emerging flux. The two loop systems collectively 
display an anenome pattern that is located below the fan surface. The flux that 
surrounds the inner spine field line of the nullpoint retains a fraction of the 
emerged twist, while the remaining twist is evacuated along the reconnected 
arcades. The nature and timing of the features which occur in the simulation 
do qualititatively reproduce those observed by XRT in the particular event 
studied in this paper. Moreover, the two-step reconnection process suggests a 
new consistent and generic model for the formation of anemone regions in the 
solar corona.

\end{abstract}


\section{Introduction}
\label{sec:int}
Various dynamic phenomena in the solar corona, as for example soft
X-ray jets and specific flares, have been associated with magnetic 
reconnection occurring in a three-dimensional (3D) magnetic nullpoint 
topology consisting of a dome-like fan separatrix surface located 
below the nullpoint and a spine field line above it \citep[e.g.][]
{lau90,ant98,pri02}. Observational evidence of 3D nullpoint topologies 
in the corona is provided by, e.g., ``saddle-like'' loop structures 
\citep{fil99}, ``ellipsoidal flare ribbons'' \citep{mas09}, and 
``anemone'' active regions within coronal holes \citep[][]{shi92}. 
The latter are characterized by a full or partial ring of radially 
aligned bright loops which connect the opposite polarities of the 
region and the surrounding coronal hole \citep[e.g.][]{asa08}. Anemone 
active regions are often associated with soft X-ray jets \citep[e.g.]
[]{shi94}, which are a strong indication of 3D nullpoint reconnection 
occurring in the corona. Further evidence for coronal nullpoint 
topologies comes from subphotospheric source models of the coronal 
magnetic field \citep{dem93}, and from potential and linear force-free 
field extrapolations of flare and jet regions \citep[e.g.][]{aul00,
fle01,uga07,mor08}. Consequently, 3D fan-spine configurations are 
increasingly used as the initial condition in numerical simulations 
of 3D reconnection, jets, and flares \citep{pon07,par09,mas09}. The 
reconnection is triggered by boundary motions in these simulations.

The {\em formation} of 3D fan-spine configurations in the corona, 
however, has not yet been studied in much detail. In potential or 
near-potential magnetic fields, a 3D nullpoint configuration with 
a fan and a spine naturally occurs if a magnetic bipole is embedded 
into one polarity region of a large-scale bipolar ambient field 
\citep[e.g.][]{ant98}. Therefore it is expected to form when magnetic 
flux emerges into regions of essentially unipolar fields, such as a 
coronal hole. The fan-spine configuration then results from the 
relaxation of the coronal field after its reconnection with the 
emerging flux. 2D numerical simulations of coronal soft X-ray jets, 
inspired by the flux emergence model by \cite{hey77}, have 
demonstrated that reconnection between emerging bipolar flux and a 
vertical or oblique coronal field yields the formation of hot loops 
connecting the ambient field with the opposite polarity flux of the 
emerging bipole \citep[e.g.][]{yok96,nis08}.  

\cite{mor08} recently performed a 3D simulation of the emergence of a 
twisted flux tube into an oblique unipolar coronal field. As in the 2D 
cases, they found the launch of a jet and the formation of a growing 
system of hot reconnected loops connecting the ambient field with the
emerging flux of opposite polarity. The resulting fan surface extends 
on one side of the emerging region, while on the other side it consists 
of non-reconnected emerged loops only. The latter are not strongly 
heated and would hence unlikely be seen in soft X-ray observations,
therefore only one half of an anemone loop pattern should be visible. 

\cite{par09} used an alternative approach to produce a jet, by starting 
from a 3D nullpoint topology and driving the jet by reconnection between 
open and closed field lines, after the latter have been significantly 
twisted by line-tied boundary motions. One outcome of this calculation 
is that, as a result of reconnection of twisted fields, the nullpoint 
moved around the axis of the spine, thus 
allowing reconnection of field lines from all sides of the fan. While 
this evolution may allow for the brightening of the whole fan in soft 
X-rays, it still does not explain how the fan-spine topology was formed 
in the first place. 

These models, although nicely reproducing coronal jets, their associated 
inverse-\textsf{Y} ``Eiffel tower'' shape, and the field line geometry 
obtained from a linear force-free field extrapolation of a jet region 
\citep{mor08}, do therefore not yet provide a satisfying scenario for the 
formation of anemone active regions. As mentioned above, these regions 
typically show a ring of bright loops below the jet that is reminiscent 
of a fan dome, which in some cases might well extend the area of emerged 
flux.

A two-dimensional effect may also play a role in the brightening of both 
sides of fan surfaces which form during flux emergence. This effect is 
the reconnection of emerging field lines back and forth with the ambient 
fields on both sides of the emerging flux. This is exactly what happens 
in the 2.5D simulation of \cite{mur09}, where an ``oscillatory reconnection'' 
pattern \citep{cra91} occurs, which the authors attributed to thermal 
pressure effects around the reconnection layer. This process was also found 
in 2.5D simulations of a quadrupolar closed field configuration, being driven 
by a non line-tied chromospheric ad-hoc monotonic force \citep{kar98}. 
Reconnection back and forth was there attributed to coronal relaxation, as a 
response to an ``overshoot'' due to a chromospheric driving which was faster 
in the simulation than on the real Sun. In the simulation of \cite{mur09}, 
the photospheric emergence velocities were small compared to the coronal 
Alfv\'en speeds (Murray, private communication, 2009), so that such an ``overshoot'' 
probably did not occur. However, the existence of (non-wave driven) oscillatory 
coronal reconnection in fully 3D configurations is yet to be established.

An interesting observational feature of coronal jets is their frequent 
transverse oscillation \citep{cir07}. Even though 2D oscillatory reconnection 
could account for such perturbations \citep{mur09}, they could also be caused 
by non-steady reconnection in a turbulent current sheet, where magnetic islands 
are gradually formed and destroyed \citep[e.g.][]{yok96,kli00,arc06}, as well 
as by upward propagating Alfv\'en waves being launched from the reconnection 
point \citep{cir07,mor08}, or by the propagation of a torsional Alfv\'en wave 
resulting from the reconnection of kinking twisted field lines with their ambient 
field \citep{par09,fil09}. The latter mechanism provides an explanation also for 
the frequently observed helical patterns traveling along jets 
\citep{shi92a,pat08,nis09}.
 
In this paper, we propose a fully 3D two-step reconnection model for 
the formation of broad fan-spine configurations in the corona. The 
model results from a zero-$\beta$ line-tied MHD simulation, in which 
the evolution of the coronal magnetic field is driven by twisted flux 
emergence prescribed at the photospheric boundary. The simulation was 
initially developed for the interpretation of a puzzling event observed 
at the solar limb by the {\it Hinode} X-ray telescope (XRT). In this 
event, two distinct small coronal loop systems developed one after the 
other beside the edge of a prominence cavity, the first one apparently 
``feeding'' the second one, while a swaying jet-like brightening was 
propagating along the cavity edge. 
Our model does not only reproduce the shape and timing of the main features 
observed in this particular event, but also accounts for the formation of 
full (and not half) anemone bright features, as generally observed in soft 
X-rays within coronal holes. It finds (in accordance with some past 
observations and models of coronal jets) that a large fraction of the 
emerged magnetic twist reconnects and is evacuated upward in the form of 
a torsional Alfv\'en wave. It furthermore shows that nullpoint reconnection 
can be accompanied by slipping reconnection \citep[e.g.][]{aul06}, which is 
supported by apparently slipping cavity loops observed by XRT in the event. 
It finally predicts that a fraction of the twist eventually remains around 
the inner spine beneath the fan/anemone surface, which therefore does not 
fully relax to a potential field configuration, even though it looks 
potential at first order.

\begin{figure}[t]
\centering
\includegraphics[width=1.0\linewidth]{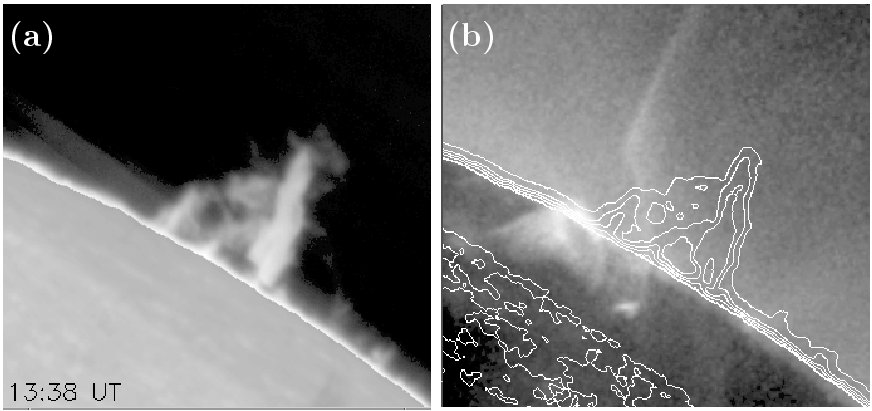}
\caption
{
{\bf (a):} MSDP $H_{\alpha}$ observation of the prominence on 2007 April 24, 
at 13:38 UT.
{\bf (b):} XRT observation of the bright loop systems on 2007 April 24, at 
18:35 UT (compare with Fig.~\ref{fig:xrt}), overlaid with $H_{\alpha}$ 
intensity contours of (a).
}
\label{fig:prom}
\end{figure}

\begin{figure*}[ht]
\centering
\includegraphics[width=0.97\linewidth]{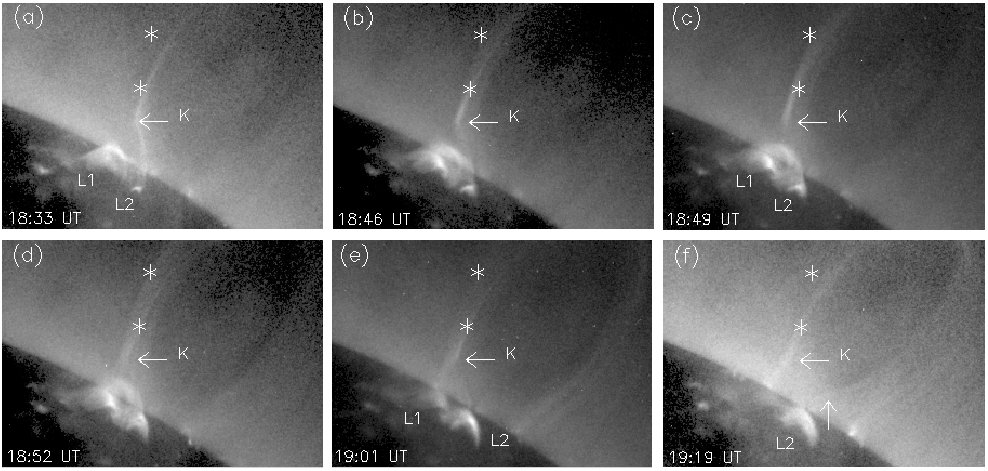}
\caption
{
Snapshots from the XRT observations ($300^{\prime\prime} \times 
212^{\prime\prime}$) at different times. The bright loop systems 
are denoted by L1, L2, and K. Asterisks and horizontal arrows are 
plotted at the same positions in all images to outline the oscillatory 
transverse motion of K and the displacement of its elbow. The vertical 
arrow points at loops which are seen to ``slip'' toward the prominence 
at the later phase of the evolution. See the text for details. (An mpeg 
animation of this figure is available in the online journal.) The 
animation uses a different intensity scaling, which outlines the cavity 
loops in more clarity.
}
\label{fig:xrt}
\end{figure*}

\section{Observations}
\label{sec:obs}
\subsection{Instrumentation}
\label{subsec:obs_ins}
The observations presented here were obtained during a coordinated campaign of 
prominence observations, involving several space and ground-based instruments. 
They were performed in the JOP 178 frame on 2007 April 23--29, during the first 
SUMER-{\it Hinode} observing campaign. JOP 178 has been running successfully many 
times in the past (see \url{http://bass2000.bagn.obs-mip.fr/jop178/}). A 
prominence surrounded by a cavity on the west limb, at S30 degree, was extensively 
studied during the campaign \citep[e.g.][]{hei08}. Fig.~\ref{fig:prom} shows an 
$\mathrm{H}_{\alpha}$ observation of the prominence obtained by the Multi channel 
subtractive double pass spectrograph (MSDP) operating at the Solar Tower of Meudon, 
as well as an overlay of the corresponding intensity contours with the XRT 
observations described below. The prominence was located in a quiet Sun region, 
apparently along the polarity inversion line (PIL) of an extended bipolar area of 
weak magnetic field, without visible strong field concentrations. A corresponding 
filament channel could be observed days before in EUV. Here we focus on XRT 
observations of a small dynamic event which took place at the edge of the cavity 
on 24 April.

The {\it Hinode} mission has been operating since 2006 October \citep{kos07}. 
XRT is a high resolution grazing incidence telescope which consists of X-ray 
and visible light optics and a 2k x 2k CCD camera. A set of filters and a 
broad range of exposure times enable the telescope to observe hot plasma in 
the range 10$^5$--10$^8$ K \citep[for more details see][]{gol07}. The 
observations presented here were obtained with a resolution of $512 \times 512$ 
pixels, each pixel having a size of $1.03'' \times 1.03''$. The filter combination 
used was Al\_poly/Open. The exposure time was $8.19$ s or $16.38$ s, the cadence 
was $60$ s. 

\subsection{Event description}
\label{subsec:obs_eve}
The brightening of several loop systems at the edge of the prominence 
cavity was observed by XRT for about two hours on 2007 April 24, 
between about 18:30 and 20:30 UT. The dynamic evolution described in 
the following lasted about 45 minutes, ending at 19:15 UT. 

Fig.~\ref{fig:xrt} shows snapshots outlining the main features of the 
evolution. At about 18:30 UT, two bright loop systems become visible, 
one set of small closed loops at the left cavity edge (denoted as L1), 
and one large loop arching over the prominence (denoted as K), outlining 
the cavity edge (Fig.~\ref{fig:xrt}a). K exhibits an elbow (indicated 
by the horizontal arrow), apparently located above L1. A very small loop 
system is observed at the apparent footpoint of K (denoted as L2 in 
Fig.~\ref{fig:xrt}a). In the following minutes, L1 continuously grows 
and the elbow slowly moves to the left. At about 18:39 UT, a brightening 
begins to propagate along K, starting from below the elbow.   

At about 18:45 UT, the evolution becomes more dynamic: L1 seems to 
expand toward L2, loops apparently connecting L1 and L2 become visible, 
and L2 starts to grow (Fig.~\ref{fig:xrt}b,c). At the same time, 
the propagating brightening has arrived at the elbow and now moves 
further along K, exhibiting a jet-like appearance. The upper part of K 
moves leftward (see asterisks) and after a while the elbow cannot be 
observed anymore. Shortly after, the part of K on the opposite side of 
the prominence starts to brighten weakly, indicating that the loop has 
been filled with hot plasma, ejected from below the elbow (the 
propagation of the jet-like brightening along K is better visible in 
the online movie accompanying Fig.~\ref{fig:xrt}). 

At 18:52 UT, L1 and L2 are roughly of the same size. They appear to have 
a common footpoint region and they collectively exhibit an anemone-like 
shape (Fig.~\ref{fig:xrt}d,e). About this time, the upper part of K 
moves back toward the right. At 19:01 UT L1 has started to fade away, and 
L2 has stopped growing (Fig.~\ref{fig:xrt}e). From about 19:10 UT on, L1 
is no longer visible. At 19:19 UT, the upper part of K has returned to its 
initial position (Fig.~\ref{fig:xrt}f). The transverse oscillatory motion 
of K is suggestive of an Alfv\'en wave traveling along it \citep{cir07}. 
Between about 19:00 UT and 19:20 UT, a successive leftward displacement 
of loops starting from the footpoint of K on the opposite edge of the 
cavity is observed (see the vertical arrow in Fig.~\ref{fig:xrt}f). After 
that, L2 slowly fades away in about an hour and no further significant 
dynamic evolution is observed.

The observations suggest that the dynamic evolution was caused by the 
interaction of newly emerging flux with the arcade-like field overlying 
the prominence. The inspection of SoHO/MDI magnetogram data from the days 
leading up to the event did not reveal significant long-lived bipolar field 
concentrations in the vicinity of the prominence, which would have indicated 
the presence of a 3D nullpoint topology before the event \citep[as in the 
simulation of][]{par09}. Since the prominence-cavity system was located at 
the limb during our event, direct observations of emerging flux are not 
available. However, the brightening of the elbowed loop system K, as well 
as the jet-like brightening propagating along it, can be understood by means 
of models of magnetic reconnection between a small emerging bipole and a 
predominantly vertical coronal field (see Sect.~\ref{sec:int}). This suggests 
that the loop system L1 was not outlining emerging fields, but rather a 
reconnected arcade that formed along with the elbow in K. The fact that L1
has a significant height when it becomes visible supports this interpretation.
Still, the transverse oscillation of K, the growth of the loop system L2, 
and the ``slipping''-like motion of loops at the opposite edge of the 
cavity cannot be understood straightforwardly within this scenario.

We note that the observed dynamics did not seem to have a noticeable 
effect on the stability of the prominence-cavity system. The prominence 
was still observed on April 25 \citep{hei08} and on April 26 by the Meudon 
Solar Tower. It hence appears that this is a case where emerging flux in 
the vicinity of a filament or prominence does not result in the eruption 
of the latter \citep[see, e.g.,][]{fey95}. 

\section{Numerical simulation}
\label{sec:simu}
In order to understand the full dynamics observed, we perform a 3D MHD 
simulation of the interaction of emerging flux with an arcade-like 
potential field overlying a coronal flux rope. The choice of such a 
coronal topology is supported by the presence of the prominence-cavity 
system. Magnetic flux ropes have been successfully used to model 
prominences and cavities \citep[e.g.][]{low95,aul98,van04,gib06a}. We 
use the analytical flux rope model by \citet[][hereafter TD]{tit99} as 
the initial condition in the simulation. The model consists of a toroidal 
current ring with major radius $R$ and minor radius $a$, which is partly 
submerged below a ``photosphere'' and is held in equilibrium by an external 
potential field created by two subphotospheric magnetic charges $\pm q$, 
which are placed at the axis of the torus, at distances $\pm L$ from the 
current ring. The coronal part of the model yields an arched, line-tied, 
and twisted flux rope which is embedded into an arcade-like potential field 
(see Fig.~2 in TD). The depth $d$ of the torus axis (and hence of the magnetic
charges) below the photospheric layer determines the compactness and strength 
of the magnetic flux distribution in the photospheric plane. Here we choose 
a relatively large depth (see below), in order to account for the observed 
extended area of weak field above which the prominence was located (see 
Sect.~\ref{subsec:obs_ins}).

Previous simulations \citep[][]{toe04,toe05,toe07,sch08} and analytical 
calculations \citep{ise07} have shown that the TD flux rope can be subject 
to the ideal MHD helical kink and torus instabilities. Therefore we use a 
weakly twisted rope here, with the field lines winding on average only once 
about the rope axis (in a right-handed sense), and we choose the potential 
field such that the rope is stable with respect to the torus instability 
\citep{bat78,kli06}.  

As in these previous simulations we integrate the $\beta=0$ compressible 
ideal MHD equations: 
\begin{eqnarray}
\partial_t\rho&=&
                 -\nabla\cdot(\rho\,\boldsymbol{u})\,,    \label{eq_rho}\\
\rho\,\partial_{t}\boldsymbol{u}&=&
      -\rho\,(\,\boldsymbol{u}\cdot\nabla\,)\,\boldsymbol{u}
      +\boldsymbol{j}\mbox{\boldmath$\times$}\boldsymbol{B} 
      +\nabla\,\cdot\bf{T}\,,
                                                          \label{eq_mot}\\
\partial_{t}\boldsymbol{B}&=& 
    \nabla\mbox{\boldmath$\times$}(\,\boldsymbol{u}\mbox{\boldmath$\times$}
    \boldsymbol{B}\,)\,,                                  \label{eq_ind}
\end{eqnarray}
where $\boldsymbol{B}$, $\boldsymbol{u}$, and $\rho$ are the magnetic
field, velocity, and mass density, respectively. The current density 
is given by 
$\boldsymbol{j}=\mu_0^{\,-1}\,\nabla\mbox{\boldmath$\times$}\boldsymbol{B}$. 
The term {\bf T} is the viscous stress tensor, included to improve 
the numerical stability \citep{toe03}. We neglect thermal pressure 
and gravity since we are interested in the qualitative evolution of 
the magnetic field only.
   
The MHD equations are normalized by quantities derived from a characteristic 
length, taken here to be the initial apex height of the TD flux rope axis
above the photospheric plane, $R-d$, the initial magnetic field strength, 
$B_0$, and Alfv\'en velocity, $v_{a0}$, at the apex of the axis, and derived 
quantities. We use a nonuniform cartesian grid of size $[-4,4] \times [-5,5] 
\times [0,8]$, resolved by $261 \times 301 \times 208$ grid points (including 
two layers of ghost points in each direction which are used to implement the 
boundary conditions), with a resolution of $\Delta x = \Delta y = 2\Delta z 
= 0.02$ in the box center. The resolution is nearly constant in the subdomain 
$[-1,1] \times [-1.5,1.5] \times [0,1]$, and increases exponentially toward 
the boundaries, to maximum values $\Delta x_{\mathrm{max}}=0.14$, 
$\Delta y_{\mathrm{max}}=0.10$, and $\Delta z_{\mathrm{max}}=0.40$. The plane 
\{$z=0$\} corresponds to the photosphere.
 
\begin{figure*}[t]
\centering
\includegraphics[width=0.97\linewidth]{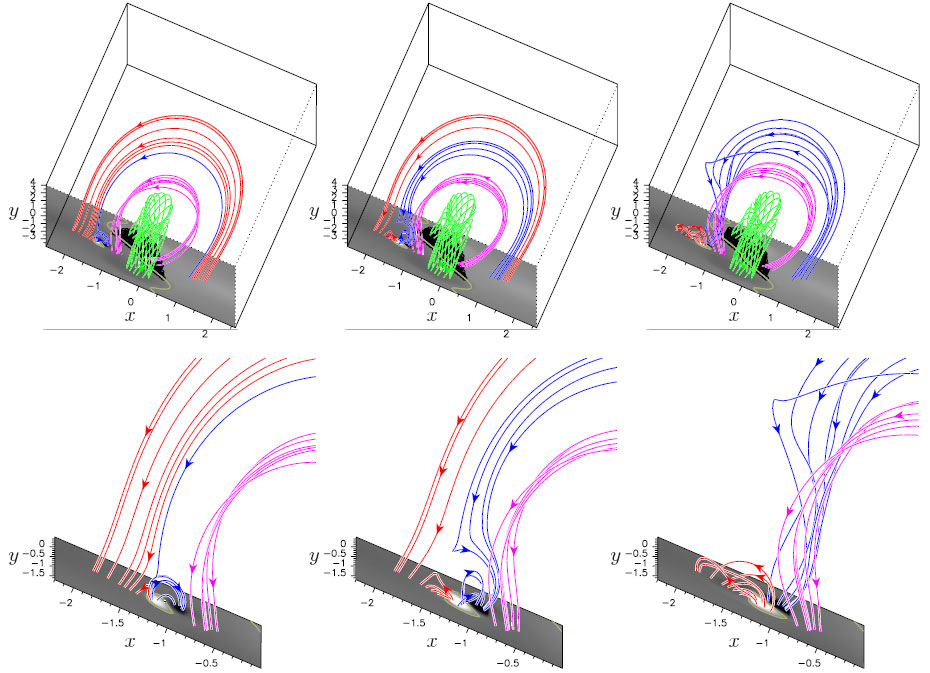}
\caption
{
Selected magnetic field lines outlining the first reconnection which occurs 
between the emerging flux rope and the outer coronal arcade. The bottom panels 
show a zoom into the emergence region. Contours of $B_z>0$ ($B_z<0$) at the 
bottom plane \{$z=0$\} are shown in white (black). Polarity inversion lines (PILs)
are drawn in yellow. The negative polarity of the large-scale potential field is
located on the left-hand side of the main PIL. All field lines are calculated 
from fixed footpoints on the left-hand side of the main PIL in all panels.
From left to right: at onset of ($t=22$), during ($t=34$), and after ($t=43$) 
the reconnection. Blue field lines the negative 
polarity of the emerging flux rope, green field lines show the core of the 
prominence flux rope. Red field lines outline the outer coronal arcade at early 
times and the newly formed small loop system later on. Pink field lines show the 
inner coronal arcade. Arrowheads mark the direction of the 
magnetic field. The black circles located above the main PIL
mark field line dips, which are assumed to carry prominence material.
(An mpeg animation of this figure is available in the online journal.)
}
\label{fig:reco1}
\end{figure*}

The TD flux rope axis is oriented along the $y$ direction, with the positive 
polarity rope footpoint rooted in the half-plane \{$y>0$\} (see Fig.~\ref{fig:reco1}). 
The normalized geometrical flux rope parameters used are: $R=2.75$, $a=1.$, 
$L=2.5$, and $d=1.75$. The top left panel in Fig.~\ref{fig:reco1} essentially 
shows the initial TD configuration, except for the small parasitic bipole and 
the blue field lines on the left-hand side of the TD flux rope. 

We employ a modified two-step Lax-Wendroff method for the integration, 
and we additionally stabilize the calculation by artificial smoothing of 
all integration variables \citep[see][for details]{sat79,toe03}. The 
reconnection occurring in our simulation is due to the resulting numerical 
diffusion. The initial density distribution is 
$\rho_0(z)=2.6\,\mathrm{exp}\,(-[z+\Delta z]/1.1)$, chosen such that the 
Alfv\'en velocity, $v_a$, slowly decreases with height above the TD flux 
rope. The system is at rest at $t=0$. 

We first perform a numerical relaxation of the system for $37$ Alfv\'en times 
and reset the time to zero afterwards. We then model the emergence of a second 
twisted flux rope in the vicinity of the TD rope, following the boundary-driven 
method by \citet[][hereafter FG]{fan03}. In their model, a toroidal twisted 
flux rope is rigidly emerged from a fictitious solar interior into a coronal 
magnetic field by successively changing the boundary conditions in the 
photospheric layer of the simulation box. We refer to Fig.~1 in FG for a sketch 
of the model. In our simulation, we choose the FG torus to be about one order of 
magnitude smaller in size than the TD torus (the major and minor radius of the 
FG torus are 0.3 and 0.2, respectively), in order to account (within the limitations 
given by the finite number of grid points) for the typically large difference in 
size between quiescent filaments and small bipoles that emerge in their vicinity 
\citep[see, e.g.,][]{fey95}, which was also suggested by the relative scale-lengths 
of the coronal loops observed in the event described in Sect.~\ref{subsec:obs_eve}.

The FG flux rope is uniformly twisted along its cross-section. The twist is 
chosen right-handed, with the field lines winding $\sim 4.5$ times along the 
whole torus. We position the rope such that the emergence region is centered 
at $(x,y)=(-1.0,-0.5)$, within the large-scale negative polarity of the TD 
potential field. As the TD flux rope, it is oriented along the $y$ direction, 
but with the opposite orientation of the axial magnetic field (see 
Fig.~\ref{fig:reco1}). The magnetic field strength within the FG torus varies 
along its cross-section, being $\sim 3\,B_0$ at the outer torus surface, 
$4 B_0$ at the axis, and $\sim 13 B_0$ at the inner torus surface, for the 
parameters used in the simulation. Although the emergence is driven until the 
apex of the inner surface approaches the bottom boundary of the box (see below), 
the field strength which effectively enters the corona during the simulation 
does typically not become larger than $\sim 6 B_0$. The field strength of the 
large-scale TD field in the small volume above the emergence area is approximately 
constant, $\sim 0.7\,B_0$. Within the TD flux rope, the field strengths vary 
between $\sim 0.6$ and $1.0\,B_0$. We discuss the rationale for our choice of 
the orientation and strength of the magnetic field within the FG flux rope in 
Sect.~\ref{sec:dis}.  
 
The boundary-driven emergence is imposed in the layer \{$z=-\Delta z$\}. 
Within the emergence area in this layer, we overwrite the pre-existing 
TD field by the respective FG flux rope field, and we set the vertical 
velocity equal to the respective driving velocity, while keeping the 
horizontal velocities at zero. Outside this area, the TD field and the 
density in \{$z=-\Delta z$\} are kept at their initial values, and the 
velocities are set to zero, at all times. These settings lead to 
significant jumps in strength and orientation of the magnetic field (i.e. 
to the formation of large values of $\nabla \cdot \bf{B}$) at the interface 
between the TD and FG fields at and close to the bottom plane. Since our 
code does not conserve $\nabla \cdot \bf{B}=0$ to rounding error, we use 
a diffusive $\nabla \cdot \bf{B}$ cleaner \citep[][]{kep03}, as well as 
Powell's source term method \citep[][]{gom94}, to minimize unphysical 
effects resulting from $\nabla \cdot \bf{B}$ errors. Furthermore, our 
overspecified boundary conditions (see above) trigger spurious oscillations, 
which after some time lead to numerical instabilities close to the bottom 
plane, in particular at the interface between the TD and FG fields. In order 
to prevent these instabilities, we apply an enhanced smoothing of all 
variables close to the boundary \citep[as in][]{toe03}, and we set the 
Lorentz force densities at \{${z=0}$\} to zero at all times. We find 
that these settings result in the formation of a rising twisted flux tube
above the emergence area, as desired.

The emergence is driven quasi-statically with a maximum velocity of 
$0.01 v_{a0}$. The driving velocity is linearly increased and decreased 
for $10$ Alfv\'en times before and after the main emergence phase (which 
lasts for $30$ Alfv\'en times), respectively. The emergence is stopped at 
$t=50$, when the apex of the inner surface of the FG torus has reached
the layer \{${z=-\Delta z}$\}. The total twist transported into the corona 
by the emerged section of the FG torus corresponds to $\sim 1.7$ field 
line turns. 

\begin{figure*}[t]
\centering
\includegraphics[width=0.97\linewidth]{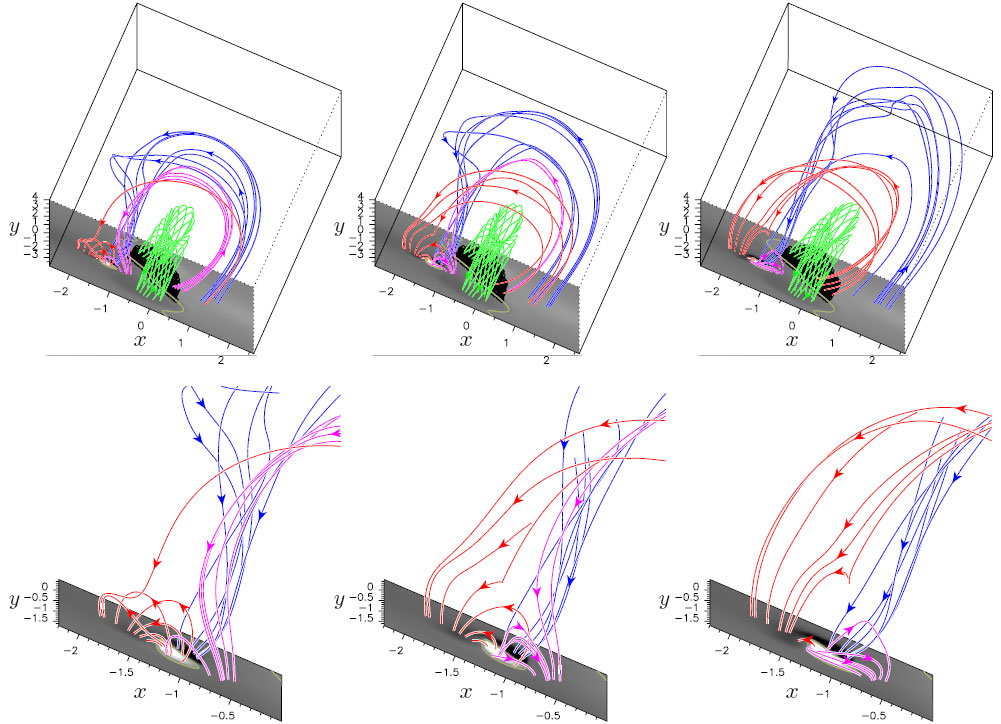}
\caption
{
Snapshots of the configuration outlining the second reconnection, which 
occurs between the newly formed small loop system and the inner coronal 
arcade. Field lines, contours, and black circles are as in 
Fig.~\ref{fig:reco1}. The bottom panels show a zoom into the emergence 
region. From left to right: at onset of ($t=46$) and during ($t=49$, $57$) 
the reconnection. Pink field lines show the inner arcade initially, and 
the second newly formed loop system later on. The propagation of an 
Alfv\'en wave along the blue field lines and the slipping motion of the 
footpoints of the red field lines on the right-hand side of the prominence 
flux rope are visible. (An mpeg animation of this figure is available in 
the online journal.)
}
\label{fig:reco2}
\end{figure*}

\begin{figure*}[t]
\centering
\includegraphics[width=0.95\linewidth]{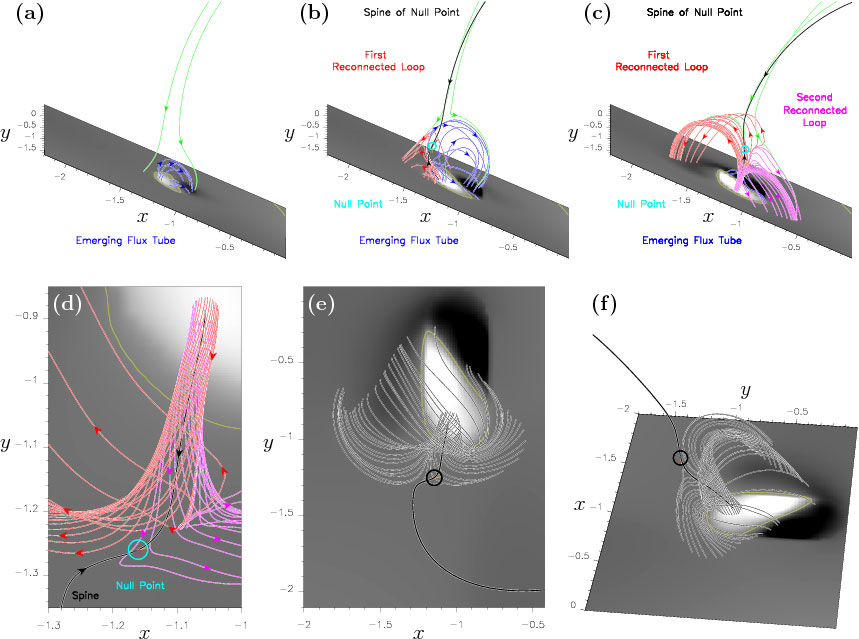}
\caption
{
{\bf (a--c):} Side view on the flux emergence region in the simulation at 
$t=17$, $34$, and $51$. The contours at the bottom plane are as in 
Figs.~\ref{fig:reco1} and \ref{fig:reco2}. Selected magnetic field lines 
are shown in order to outline the topological evolution.  
{\bf (d):} Top view on the center of the anemone-shaped loop pattern shown 
in (c), outlining the twist of the field lines.
{\bf (e)} and {\bf (f):} Top and side view on the emergence region at $t=51$,
showing field lines starting from a circle around the spine (black line) 
with twice the radius as used in (c) and (d). These field lines approximately 
outline the fan surface. The respective positions of the nullpoint are 
indicated by circles. 
}
\label{fig:topo}
\end{figure*}

\section{Simulation results and comparison with XRT observations}
\label{sec:simu_result}
In this section, we describe our simulation results and compare them with 
the XRT observations described in Sect.~\ref{subsec:obs_eve}. The interaction 
between the emerging FG flux rope and the pre-existing TD coronal field 
results in two distinct reconnection phases, which are described in 
Sects.~\ref{subsec:firststep} and \ref{subsec:secondstep}, respectively. 
In Sect.~\ref{subsec:geom}, we discuss the magnetic field geometry resulting 
from the reconnections, and in Sect.~\ref{subsec:promresponse} we describe the 
response of the TD flux rope and its surrounding arcade to the dynamics
accompanying the reconnection. In order to emphasize that the evolution
found in the simulation does not rely on the specific flux rope models
used, we refer to the FG rope as ``emerging flux rope'', and to the TD 
rope as ``prominence flux rope'' throughout this section.

Figs.~\ref{fig:reco1}--\ref{fig:topo} display magnetic field lines which 
outline the main features of the magnetic field evolution in the coronal 
domain. Figs.~\ref{fig:reco1} and \ref{fig:reco2} focus on the dynamics, 
showing the first and second reconnection phase, respectively. 
Fig.~\ref{fig:topo} shows the evolution of the magnetic topology. The blue 
field lines in Figs.~\ref{fig:reco1}--\ref{fig:topo} are integrated starting 
from the negative polarity of the emerging flux rope, green field lines 
in Figs.~\ref{fig:reco1} and \ref{fig:reco2} outline the core of the 
prominence flux rope (black circles mark field line dips). Red (pink) field 
lines in Figs.~\ref{fig:reco1} and \ref{fig:reco2} show the outer (inner) 
arcade overlying the prominence flux rope initially, and reconnected field 
lines later on. Note that all field lines are calculated from the same 
positions on the left-hand side of the prominence flux rope in all panels.

\subsection{First step: formation of one half of the anemone and jet acceleration}
\label{subsec:firststep}
As the emerging flux rope (closed blue field lines in the bottom 
panels of Fig.~\ref{fig:reco1}) enters the coronal domain, it
starts to expand and a current sheet forms above the outer edge 
of its positive polarity, at the interface of the rope and the 
outer coronal arcade that surrounds the prominence flux rope. 
Since the outermost emerging fields and the outer arcade fields 
are oppositely directed at the location of the current sheet, the 
two flux systems readily start to reconnect, forming a new small 
loop system below the current sheet and strongly bent, elbow-shaped 
field lines above it (cusp-shaped red field lines and blue field 
lines in the central panels of Fig.~\ref{fig:reco1}, respectively). 
Note that the not yet reconnected part of the emerging flux rope 
continues to expand in the corona after the reconnection has started.
The shape of the reconnected field lines agrees very well with the 
shape of the bright loops observed by XRT in the early phase of the 
event described in Sect.~\ref{subsec:obs_eve} (see L1 and K in 
Fig.~\ref{fig:xrt}a), which indicates that the observed loops have 
been formed in an analogous reconnection process. 

The field line shapes are a signature of a fan-spine configuration in 
a 3D nullpoint topology (see also Fig.~\ref{fig:topo}b). A magnetic 
nullpoint is indeed formed within the current sheet in the simulation, 
right after the reconnection has started. It forms as the system tends 
to relax to a lower energy state\footnote{The lowest possible energy 
state for any photospheric magnetic field distribution which develops 
during the emergence process would locally (i.e. above the emergence
region) correspond to a potential field that has to contain one single 
nullpoint, owing to the presence of a closed PIL embedded in a region 
of nearly vertical field \citep[e.g.][see Introduction]{ant98}.}. The 
reconnection continues as time evolves, thus the size of the reconnected 
red loop system increases (Fig.~\ref{fig:reco1}), in agreement with the 
observed growth of L1 (Fig.~\ref{fig:xrt}a,b). The ongoing expansion of 
the emerging flux rope initially slowly pushes the nullpoint away from 
the prominence flux rope, in agreement with the observed leftward 
displacement of the elbow indicated by the horizontal arrow in 
Fig.~\ref{fig:xrt}.

\begin{figure}[t]
\centering
\includegraphics[width=0.70\linewidth,clip=]{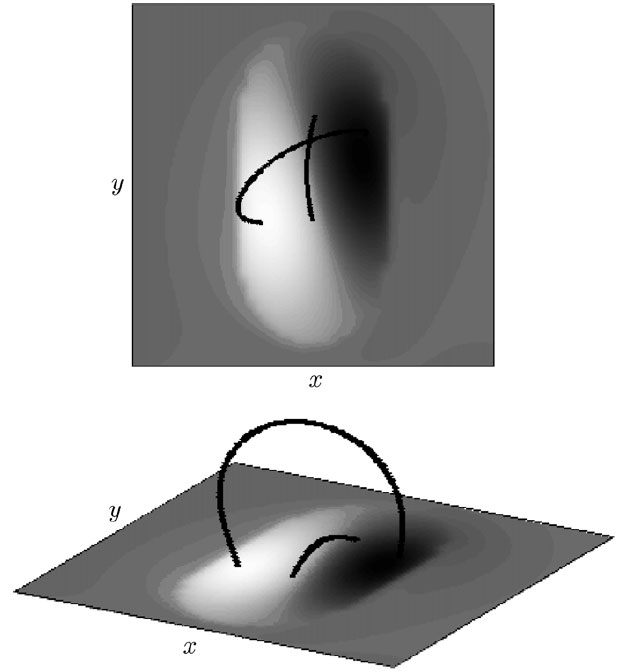}
\caption
{
Top and oblique view on two field lines within the emerging flux rope 
at $t=31$, outlining the change of field line inclination during the 
emergence. See the text for details.} 
\label{fig:twist}
\end{figure}

As the emergence continues, the emerging flux rope field lines become 
increasingly sheared with respect to the surrounding coronal arcade. 
This is due to the fact that the twist within the FG flux rope is nearly 
uniform along its cross-section (as in the well-known Gold-Hoyle model). 
While field lines far away from the flux rope axis are strongly inclined 
with respect to the axis, field lines close to the axis are almost aligned 
with it. Hence, as the flux rope emerges, its outer field lines resemble 
a nearly potential coronal arcade that is oriented almost orthogonal to 
the local PIL, whereas its inner field lines (i.e. those close to its axis) 
resemble a small sheared coronal arcade (see Fig.~\ref{fig:twist}). The 
first flux rope field lines which reconnect with the large-scale coronal 
arcade are thus almost unsheared with respect to the arcade (see the 
bottom left panel in Fig.~\ref{fig:reco1}). As the evolution continues, 
progressively more sheared loops are reconnected. As a result, the system 
of new reconnected (red) field lines eventually develops a shear distribution 
that is opposite to the one of the emerging flux rope: the field lines are 
sheared at the edges of the system, and almost unsheared close to the local 
PIL (see the bottom right panel in Fig.~\ref{fig:reco1} and the corresponding 
online animation).

The reconnection does not only yield the transfer of twist from the emerging 
flux rope into the newly formed red loop system. Part of the flux rope twist 
is also transferred into the lower parts of the reconnected overlying blue 
field lines that are now rooted in the negative polarity of the rope. Since 
the upper parts of these field lines are nearly potential, whereas their 
lower parts experience a sudden injection of twist, they are far from being 
force-free. Their relaxation is ensured through the launch of a torsional 
Alfv\'en wave which travels from low altitudes all along the arcade (see the 
evolution of the blue field lines in Figs.~\ref{fig:reco1} and \ref{fig:reco2}). 

Such reconnection between twisted and untwisted coronal fields has been suggested 
by several authors as a driving mechanism for jets \citep[see, e.g.,][]{sch95,shi97}. 
If the reconnection is sufficiently impulsive, it can launch a shear (in 2.5D) or 
torsional (in 3D) Alfv\'en wave, which can accelerate the plasma upward, as shown 
in numerical simulations by \cite{yok99} and \cite{par09}, respectively. The 
impulsive nature and large wavenumber of the wave in our simulation is a priori 
not expected, since the transition from nearly unsheared to highly sheared emerging 
flux rope fields lines involved in the reconnection is continuous. Also, our code 
does not incorporate any time-varying resistivity. Therefore, the impulsive launch 
of the wave must result from some perturbation of the system which yields a strong 
increase of the reconnection rate. Indeed, we find a strong expansion of the not 
yet reconnected central part of the emerging flux rope at $t \approx 40$ (i.e. 
between the stages shown in the middle and right panels of Fig.~\ref{fig:reco1}; 
see the corresponding online animation). At this time, a flux rope twist of 
$\sim 1.5$ turns has entered the corona, indicating that this sudden increase 
in expansion might be related to the onset of a kink instability 
\citep[as in][]{fan03,fan07}. The reconnection-driven torsional Alfv\'en wave 
in our simulation suggests an explanation for the jet-like brightening traveling 
along the cavity loops observed by XRT in our event, as well as for the observed 
transverse oscillation of the upper parts of the cavity loops 
(see Sect.~\ref{subsec:obs_eve}). 

The transverse deformation of the blue field lines in our simulation during 
the passage of the wave (Fig.~\ref{fig:reco2}) is, however, obviously much 
larger than what is observed in our event and typically in coronal jet-like 
events \citep{cir07}. This might simply be due to the fact that, although we 
have chosen the emerging flux rope to be as small as possible as compared to 
the prominence flux rope (see Sect.~\ref{sec:simu}), the difference in size 
between the emerging flux and the prominence-cavity system might still be 
significantly larger. Also, to some extent the unrealistic reconnection time 
scales in our simulation might play a role. They are mostly constrained by the 
intrinsic diffusivity of the numerical scheme and by the prescribed magnetic 
field smoothing, and do neither correspond to fast reconnection nor to the 
seminal Sweet-Parker regime. Still, the qualitative agreement which we find 
here with other coronal jets simulations performed with different codes (see 
Sect.~\ref{sec:int}) suggests that we can believe in the overall mechanism for 
jet acceleration which our simulation finds. 

Note that the nullpoint-related elbow in the lower part of the blue field 
lines apparently disappears during the evolution in our simulation (as it 
does in the observation too; see Fig.~\ref{fig:xrt}). In the simulation, 
this is merely due to projection effects and the motion of the nullpoint. 
From a different viewing angle, an elbow at low heights remains visible. 

Up to this point, the evolution is as expected from the classical model 
for coronal jets and previous 2.5D and 3D simulations of this model (see 
Sect.~\ref{sec:int} and Fig.~\ref{fig:topo}a,b). It results in the 
formation of a 3D nullpoint topology, but actually half of its fan surface 
still consists of not reconnected emerged field lines, so that only 
the other half is expected to brighten in soft X-rays. In other words, 
only a {\em half} anemone has been formed at this stage in the simulation. 

\subsection{Second step: formation of the second half of the anemone}
\label{subsec:secondstep}
As the reconnection-driven transfer of twist/shear from the emerging
flux rope into the ambient coronal field progresses, the footpoints 
of the newly formed red field lines located within the positive polarity 
of the emerging flux rope are continuously displaced in the negative 
$y$ direction (see Fig.\,\ref{fig:reco1} and the online animation), 
toward inner arcade field lines which also overlie the prominence flux 
rope and are yet unaffected by the reconnection (shown in pink in 
Fig.\,\ref{fig:reco1}). Meanwhile, following the magnetic field 
reorientation at high altitude, the nullpoint (and hence the reconnection 
site) undergoes a counterclockwise horizontal rotation, from the leftmost 
edge of the emerging rope toward its center, thereby approaching the pink 
arcade field lines (see Fig.~\ref{fig:topo}). The displacement of the 
footpoints of the small red loop system corresponds to the apparent 
expansion of L1 toward L2 in the observation, which starts at about 
18:45 UT (see Sect.~\ref{subsec:obs_eve}).  

Eventually a second reconnection starts, now between the previously 
reconnected sheared red field lines and the pink arcade field lines 
(Fig.~\ref{fig:reco2}). It must, and indeed does, take place at the 
nullpoint, which has gradually moved toward the pink arcade during 
the first reconnection episode (see above). This second reconnection 
leads to the formation and growth of a second small loop system 
(cusp-shaped pink field lines in Fig.\,\ref{fig:reco2}), which 
corresponds to the growth of the observed bright loop system L2 from 
about 18:50 UT on (Fig.~\ref{fig:xrt}). The reconnected red and pink 
loop systems both have footpoints within the positive polarity of the 
emerging flux, and they collectively display an anemone-like shape 
which is significantly wider than this polarity (Fig.~\ref{fig:topo}c; 
compare also with the collective shape of L1 and L2 in 
Fig.~\ref{fig:xrt}d,e).  

In the course of the second reconnection, the sheared red loop system 
``feeds'' the newly formed small pink loop system with magnetic flux, 
hence the former shrinks while the latter grows (Fig.~\ref{fig:reco2}). 
Although, at first glance, the bottom right panel gives the impression 
that the red loop system has almost disappeared, the elbow visible in 
the rightmost red arcade field line shows that there still exist 
relatively extended, not reconnected red loop field lines at this late 
phase of the second reconnection. These field lines are rooted between 
the elbow-shaped field line and the rightmost, very flat red field line. 
The final relative size of the red and pink loop systems will depend on 
how long reconnection persists, which, in turn, will depend on a number 
of factors, as for example on the ratio of the respective magnetic fluxes 
present in the reconnecting flux systems and on the lowest state of 
magnetic energy the system can reach by means of its relaxation. A 
shrinking of L1 during the growth of L2 is also indicated by the 
observations (compare Fig.~\ref{fig:xrt}d and e). However, this 
``shrinking'' might be merely due to a relatively fast cooling of L1. 
Studying the issues of lifetime and visibility in soft X-ray of the 
reconnected loops in our simulation would require the inclusion of proper 
thermodynamics, which is beyond the scope of this paper.

\subsection{Geometrical properties of the reconnected field}
\label{subsec:geom}
The geometrical properties of these new loop systems in the simulation
are different from what has been found in the simulations mentioned in 
Sect.~\ref{sec:int}. They arise from a two-step and fully 3D transfer 
of sheared flux, first from the sheared core of the emerging flux 
rope into the new small loop system formed at its side (as the emerging 
blue field lines reconnect with ambient [red] arcade field lines that are 
anchored on the left-hand side of the emerging rope), and second from the 
very same loop system into a second generation of small loops, which form 
at the other side of the emerging flux rope (as the sheared small red 
field lines reconnect with other ambient [pink] arcade field lines). 

As a result, a {\em full} anemone forms around the parasitic polarity 
of the emerging bipole. A large fraction of the nullpoint associated 
fan now consists of once or twice reconnected field lines, and not to 
a large extent of emerged field lines as in most of the models described 
in Sect.~\ref{sec:int}. Since both sides of the fan have been formed 
through magnetic reconnection, they can a priori be both visible as hot 
loops in soft X-rays. The center of the anemone structure contains 
significant twist once it has fully formed, although the structure appears 
to be potential when viewed from some distance (Fig.~\ref{fig:topo}c). 
The twist is concentrated around the inner spine of the nullpoint (see 
Fig.~\ref{fig:topo}d), but some of it is also present along the fan. This 
twist is the remnant of magnetic shear that has not been ejected in the 
form of a torsional Alfv\'en wave along the large-scale reconnected 
arcades that overlie the prominence flux rope.

It is worth noting that, even though a fully 3D anemone has been formed, 
it still locally possesses a quasi, but not exact, translational invariance 
along the $y$ axis, i.e. along the axis of the emerging flux rope, around 
the nullpoint. This can be seen in Fig.~\ref{fig:topo}c,d: almost all 
of the red and pink field lines fan out roughly along the $x$ direction, i.e. 
perpendicular to the rope axis, even though they have been integrated from 
equidistant footpoint positions along a small circle centered at the inner 
spine of the nullpoint, within the positive polarity of the emerging flux, 
where $B_z$ is roughly constant (as shown in Fig.~\ref{fig:topo}d). This 
strong departure from axisymmetry around the nullpoint, which does not exist 
in all of the 3D nullpoint models \citep[e.g.][]{ant98,par09}, can here be 
attributed to the elongation of the emerging flux, and possibly to the 
different inclinations of the ambient field lines. The few red and pink 
field lines in Fig.~\ref{fig:topo}d which fan out from the nullpoint region 
along the positive $y$ direction, clearly show that the anemone does not 
contain a true null line, but one single nullpoint. The quasi-translational 
invariance is then due to fan-related eigenvalues of the single nullpoint, 
which have very different amplitudes \citep{lau90}. This property, which 
has already been identified in the simulation of \citet{mas09} with a 
different MHD code, may be present also in other simulations, such as those 
which let emerge a very long flux rope \citep[e.g.][]{mor08}.

Fig.~\ref{fig:topo}e and f show that, despite the strong departure from 
axisymmetry, the fan surface also extends in $y$-direction, i.e., along the 
axis of the emerging flux rope. The field lines shown in the two panels are 
located very close to the fan surface and outline a ``heart-shaped'' form 
of the fan. We expect the fan to develop a more uniform radial distribution 
as the system relaxes to a force-free field, or if less elongated parasitic 
polarities are considered \citep{ant98}.   

\subsection{Response of the prominence cavity to the anemone/jet formation}
\label{subsec:promresponse}
While the two-step reconnection described in the previous subsections
persists beside the prominence flux rope, the torsional Alfv\'en wave 
triggered during the first reconnection phase travels upward (see the 
evolution of the blue field lines in the top panels of Fig.\,\ref{fig:reco2}). 
Even though the upper part of the prominence flux rope significantly bends 
to the side while the wave passes by, the simulated ``prominence material'' 
located in the flux rope dips \citep[computed as in ][and indicated by the 
black circles in Figs.~\ref{fig:reco1} and \ref{fig:reco2}]{aul98,van04,gib06a,dud08} 
does not show a significant motion. This difference is due to the relatively 
small twist of the rope combined with its non-negligible curvature, which 
result in the occurrence of magnetic dips only far below the rope axis, 
at altitudes low enough not to be significantly affected by the wave
(see also the discussion on the strength of the perturbation caused by 
the wave in Sect.~\ref{subsec:firststep}). 

A perturbation of the cavity is observed by XRT on the right-hand side of 
the prominence in our event. Between 19:01 and 19:19 UT, after the 
formation of L1, while the jet-like brightening is still propagating 
and L2 is still growing, a continuous motion (``slipping'') of the lowermost 
parts of several loops, from the outer edge of the cavity toward the prominence, 
becomes visible (see Fig.~\ref{fig:xrt}e, Fig.~\ref{fig:xrt}f, and the 
corresponding XRT movie). Estimating their slipping velocities by following 
them individually is not straightforward, as all these loops are not very much 
contrasted with respect to the background corona. The only sure number we could 
derive is a minimum drift velocity of $35\,\mathrm{km\,s^{-1}}$ for the ensemble 
of these loops, since they all have moved by $51^{\prime\prime}$ along the solar 
limb during a time interval of 18 minutes. Still, we cannot rule out that the 
drift velocity of individual loops could be much larger. 

The very same phenomenon takes place in the simulation. The online animation of 
the simulation shows a slipping motion of the footpoints of the red large-scale 
arcade field lines on the right-hand side of the prominence flux rope (see also 
the top panels in Fig.~\ref{fig:reco2}). This slipping occurs along the footpoints 
of arcade field lines (of which some are shown by the pink field lines in the left 
panels of Fig.~\ref{fig:reco2}). It starts right after the second reconnection 
sets in and begins to exchange the connectivities between the small (and already 
once reconnected) sheared red field lines and large-scale arcade field lines. This 
slipping of field lines, after they have reconnected at the nullpoint, is not a 
numerical artifact, caused by insufficient accuracy of the field line integration 
or by a too strong smoothing of the magnetic field during the MHD simulation: the 
slipping occurs over many grid points in the large-scale positive polarity beside 
the prominence flux rope, and it does not occur outside of flux regions which 
reconnect at the nullpoint. A very mild slipping is noticeable also during the 
first reconnection, at the footpoints of the reconnected large-scale blue lines 
along which the wave travels (see the online animation). 

The explanation of this slipping is very probably the same as first put 
forward by \cite{mas09} in their nullpoint associated confined flare model: 
the asymmetry of the fan-spine configuration, manifesting as a local 
quasi-translational invariance of the magnetic field around a nullpoint,
which has one fan-related eigenvalue of very small, but finite value (see 
Sect.~\ref{subsec:secondstep} and Figs.~\ref{fig:topo}c,d), results in the 
appearance of a narrow halo of finite width around the nullpoint separatrices, 
in which field lines have strong squashing degrees, i.e. which constitutes 
quasi-separatrix layers \citep[QSLs;][]{dem96}. It has been recently shown 
that current sheet formation and diffusion at QSLs result in 
slipping/slip-running magnetic reconnection, manifesting as apparent field 
line motions at sub/super Alfv\'enic speeds \citep{aul05,aul06}. In the case 
of a 3D nullpoint topology, the presence of a QSL halo surrounding both the 
fan surface and the spine field line results in a complex pattern for magnetic 
reconnection, in which a given field line slips and slip-runs, both before and 
after its connectivity jumps at the nullpoint. To the best of our knowledge, 
the present simulation is the second one after \cite{mas09} for which this 
sequential nullpoint and slipping reconnection behavior is reported. Based on 
these simulations, the slipping loops observed by XRT in our event can be 
regarded as a new direct evidence for slipping reconnection in the corona, 
as previously observed by \cite{aul07}. 

\section{Discussion}
\label{sec:dis}
We presented a $\beta=0$ 3D MHD simulation of the interaction of a small 
emerging bipole with a large-scale arcade-like coronal field overlying a 
weakly twisted coronal flux rope. The simulation was developed in order 
to understand {\it Hinode}/XRT observations of a small limb event which 
took place at the edge of a quiet Sun prominence cavity. The event showed 
a number of puzzling features, which could satisfactorily be explained by 
the magnetic field evolution in the simulation 
(see Sect.~\ref{sec:simu_result} for details).    

The main results of the simulation can be summarized as follows:

\begin{enumerate}

\item The emergence of a twisted flux rope into one large-scale polarity 
of an arcade-like coronal field yields the formation of a fully 3D single 
nullpoint topology in the corona, consisting of a fan-spine configuration 
in which the fan surface {\em significantly extends} over the parasitic 
polarity of the emerged flux.

\item The configuration forms in a {\em two-step reconnection} process at 
the nullpoint, which yields the successive formation of two small loop 
systems below the fan surface, on opposing sides of the parasitic polarity. 
The first loop system hereby ``feeds'' the second one with magnetic flux. 
Since the two loop systems (and field lines surrounding them) have common 
footpoints in the parasitic polarity, they collectively display an 
{\em anemone} shape.
 
\item The reconnection facilitates the transfer of twist/shear from the 
emerging flux rope into the coronal arcades by means of a {\em torsional 
Alfv\'en wave} which travels along the arcades. A fraction of the twist 
remains below the fan surface, where therefore the resulting magnetic field 
is nonpotential.

\item The wave is launched by reconnection, as the expansion velocities of
the emerging fields in the corona suddenly increase, possibly due to the
onset of a ideal MHD kink instability in the not yet reconnected part of 
the emerging flux rope, after sufficient twist has entered the corona.

\item The 3D nullpoint reconnection is accompanied by {\em slipping 
reconnection} of arcade field lines on the opposite side of the 
pre-existing coronal flux rope. This can probably be explained by the 
presence of quasi-separatrix layers around the nullpoint.

\end{enumerate}

Our simulation combines dynamic and topological elements of flux emergence 
into unipolar coronal fields and coronal jet formation which have previously 
not been found altogether in a single simulation. The two-step reconnection 
process found in our simulation provides a new model for the formation of 
3D nullpoint topologies with extended fan surfaces in the corona. By means 
of reconnection between twisted and untwisted fields, the model can also 
account for coronal jet activity \citep[e.g.][]{sch95,shi97,par09}. 
Finally, since the fields below the fan surface are all formed by reconnection, 
and can therefore a priori assumed to brighten in soft X-rays, our simulation 
also suggests a mechanism for the formation of (full) anemone active regions.
The fact that some of the emerging twist remains below the fan surface in the 
final configuration supports models of jet formation which assume a twisted 
null point configuration prior to the jet \citep[e.g.][]{par09}.

Our model can, however, not account for the {\em long-lived evolution} of anemone 
active regions. Such regions are observed to consist of bright loops on both sides 
of a parasitic polarity over time-periods of days \citep[e.g.][]{asa08}, whereas 
the reconnection in our simulation is expected to produce rather short-lived 
brightenings, which would occur successively rather than simultaneously (as in the 
event described in this paper). It seems likely that the long-lived dynamics of 
anemone regions is due to continuous dynamic perturbations of the configuration 
after its formation by, for instance, ongoing flux emergence and photospheric flows. 
Our model might serve as a starting point for a numerical study of anemone region 
dynamics.

Let us now briefly discuss the robustness of our results with respect to the variation 
of various model parameters. A detailed investigation of this question will require an 
extensive parametric study, but some aspects may be discussed here.

First of all, the model relies on fully 3D effects. The most important ingredient 
is the twist brought up by the emerging flux rope. For the two-step reconnection 
as described in this paper to work, the emerging field lines must significantly 
change their inclination as they rise into the corona, such that the first 
reconnected loop system can evolve toward yet unreconnected remote ambient fields 
during the reconnection. If the twist in the emerging flux rope (more precisely: 
the change of field line inclination along the cross-section of the rope) is chosen 
too small, the first reconnection will occur (as in the flux emergence simulations 
mentioned in Sect.~\ref{sec:int}), but the second one will not set in. Our adopted 
boundary-driven ``kinematic'' flux emergence technique models the rigid emergence 
of a twisted flux rope into the corona, and does therefore not capture the dynamics 
of real flux emergence \citep[for a discussion, see][]{fan03}. In particular, it 
assures that the inclination of the emerging field lines is continuously changing. 
It would be, of course, desirable to test our model in a ``dynamic'' flux emergence 
simulation, as the one by \cite{mor08}. Here we have chosen the ``kinematic'' 
approach since, at present, ``dynamic'' flux emergence simulations do not allow to 
consider complex coronal magnetic field configurations as the one used here, which 
was chosen to model the observed prominence-cavity system.

Apart from the twist profile of the emerging flux rope, there are several other 
model parameters which had to be chosen carefully in order to match the observations
(but not necessarily to produce a 3D nullpoint configuration, which is more general). 
First, of course, the size of the emerging flux rope and its distance from the coronal 
flux rope had to be adjusted as suggested by the observations. Second, the choice of 
the magnetic field strength within the emerging rope (and therefore its magnetic flux ) 
is important. The ratio of the field strength (and flux) between the emerging rope and 
the ambient corona will influence, for instance, the amplitude of the torsional 
Alfv\'en wave and the final size of the reconnected small loops, i.e. the size of the 
anemone. In the simulation presented here, the field strength within the coronal part 
of the emerging flux rope was typically about five times larger than in the neighboring 
arcade fields throughout most of the evolution, so that the flux within the (small) 
rope was not negligible compared to the flux contained in the large-scale arcades. 
Third, the inclination of the initial coronal field might play a role, although we 
expect a similar evolution if the flux rope would be emerged into a purely vertical 
coronal field, mimicking ``open'' field lines in coronal holes. Finally, the magnetic 
orientation of the emerging rope with respect to the pre-existing coronal configuration 
plays an important role. If this quantity would be reversed in the simulation, and if 
the evolution would be seen from the same viewing angle as in Figs.~\ref{fig:reco1} and 
\ref{fig:reco2}, we expect the second reconnected (pink) loop system to form on the 
left-hand side of the first (red) one, and the blue reconnected field lines above the 
nullpoint to develop an elbow which bends away from the prominence flux rope rather 
than toward it. It would therefore be interesting to study how the system behaves if 
intermediate orientations of the emerging rope are chosen. 
 
Our results underline the importance of a precise examination of 
the magnetic topology (and of its formation) for the understanding 
of many dynamic events in the solar corona. Without a detailed 
consideration of the topology, it would have been very difficult
to understand the complex sequence of dynamic features occurring 
in the simulation and in the observed event.

\acknowledgments
We thank the referee for detailed comments and suggestions which helped
very much to improve the quality of this paper.
Financial support by the European Commission through the SOLAIRE Network
(MTRN-CT-2006-035484) and through the FP7 SOTERIA project (Grant Agreement 
n$^o$ 218816) are gratefully acknowledged. {\it Hinode} is a Japanese 
mission developed and launched by ISAS/JAXA, with NAOJ as domestic partner 
and NASA and STFC (UK) as international partners. It is operated by these 
agencies in co-operation with ESA and the NSC (Norway). LG and KKR are 
supported by NASA under contract NNM07AB07C to SAO.
We thank all the people who collaborate actively within the 
JOP 178 observations campaign, particularly T. Berger and the {\it Hinode} 
team, G. Molodij and the Meudon Solar Tower observers, and P. Mein for 
reducing the MSDP data, and T. Roudier 
and N. Labrosse for the coordination and the pointing of the observations.
TT thanks Y. Fan for several discussions on ``kinematic flux emergence''.
We also thank P. D\'emoulin, B. Kliem, and E. Pariat for very helpful
comments on this work. Our MHD calculations were done on the dual-core
quadri-Opteron computers of the Service Informatique de l'Observatoire de 
Paris (SIO).

\bibliographystyle{apj}
\bibliography{toeroek}  

\begin{thebibliography}{58}
\expandafter\ifx\csname natexlab\endcsname\relax\def\natexlab#1{#1}\fi

\bibitem[{{Antiochos}(1998)}]{ant98}
{Antiochos}, S.~K. 1998, \apjl, 502, L181

\bibitem[{{Archontis} {et~al.}(2006){Archontis}, {Galsgaard},
  {Moreno-Insertis}, \& {Hood}}]{arc06}
{Archontis}, V., {Galsgaard}, K., {Moreno-Insertis}, F., \& {Hood}, A.~W. 2006,
  \apjl, 645, L161

\bibitem[{{Asai} {et~al.}(2008){Asai}, {Shibata}, {Hara}, \& {Nitta}}]{asa08}
{Asai}, A., {Shibata}, K., {Hara}, H., \& {Nitta}, N.~V. 2008, \apj, 673, 1188

\bibitem[{{Aulanier} {et~al.}(2000){Aulanier}, {DeLuca}, {Antiochos},
  {McMullen}, \& {Golub}}]{aul00}
{Aulanier}, G., {DeLuca}, E.~E., {Antiochos}, S.~K., {McMullen}, R.~A., \&
  {Golub}, L. 2000, \apj, 540, 1126

\bibitem[{{Aulanier} \& {D\'emoulin}(1998)}]{aul98}
{Aulanier}, G., \& {D\'emoulin}, P. 1998, \aap, 329, 1125

\bibitem[{{Aulanier} {et~al.}(2005){Aulanier}, {D{\'e}moulin}, \&
  {Grappin}}]{aul05}
{Aulanier}, G., {D{\'e}moulin}, P., \& {Grappin}, R. 2005, \aap, 430, 1067

\bibitem[{{Aulanier} {et~al.}(2007){Aulanier}, {Golub}, {DeLuca}, {Cirtain},
  {Kano}, {Lundquist}, {Narukage}, {Sakao}, \& {Weber}}]{aul07}
{Aulanier}, G., {Golub}, L., {DeLuca}, E.~E., {Cirtain}, J.~W., {Kano}, R.,
  {Lundquist}, L.~L., {Narukage}, N., {Sakao}, T., \& {Weber}, M.~A. 2007,
  Science, 318, 1588

\bibitem[{{Aulanier} {et~al.}(2006){Aulanier}, {Pariat}, {D{\'e}moulin}, \&
  {Devore}}]{aul06}
{Aulanier}, G., {Pariat}, E., {D{\'e}moulin}, P., \& {Devore}, C.~R. 2006,
  \solphys, 238, 347

\bibitem[{{Bateman}(1978)}]{bat78}
{Bateman}, G. 1978, {MHD instabilities} (Cambridge, Mass., MIT Press)

\bibitem[{{Cirtain} {et~al.}(2007){Cirtain}, {Golub}, {Lundquist}, {van
  Ballegooijen}, {Savcheva}, {Shimojo}, {DeLuca}, {Tsuneta}, {Sakao}, {Reeves},
  {Weber}, {Kano}, {Narukage}, \& {Shibasaki}}]{cir07}
{Cirtain}, J.~W., {Golub}, L., {Lundquist}, L., {van Ballegooijen}, A.,
  {Savcheva}, A., {Shimojo}, M., {DeLuca}, E., {Tsuneta}, S., {Sakao}, T.,
  {Reeves}, K., {Weber}, M., {Kano}, R., {Narukage}, N., \& {Shibasaki}, K.
  2007, Science, 318, 1580

\bibitem[{{Craig} \& {McClymont}(1991)}]{cra91}
{Craig}, I.~J.~D., \& {McClymont}, A.~N. 1991, \apjl, 371, L41

\bibitem[{{D{\'e}moulin} {et~al.}(1996){D{\'e}moulin}, {H{\'e}noux}, {Priest},
  \& {Mandrini}}]{dem96}
{D{\'e}moulin}, P., {H{\'e}noux}, J.~C., {Priest}, E.~R., \& {Mandrini}, C.~H.
  1996, \aap, 308, 643

\bibitem[{{D{\'e}moulin} {et~al.}(1993){D{\'e}moulin}, {van Driel-Gesztelyi},
  {Schmieder}, {Hemoux}, {Csepura}, \& {Hagyard}}]{dem93}
{D{\'e}moulin}, P., {van Driel-Gesztelyi}, L., {Schmieder}, B., {Hemoux},
  J.~C., {Csepura}, G., \& {Hagyard}, M.~J. 1993, \aap, 271, 292

\bibitem[{{Dud{\'{\i}}k} {et~al.}(2008){Dud{\'{\i}}k}, {Aulanier}, {Schmieder},
  {Bommier}, \& {Roudier}}]{dud08}
{Dud{\'{\i}}k}, J., {Aulanier}, G., {Schmieder}, B., {Bommier}, V., \&
  {Roudier}, T. 2008, \solphys, 248, 29

\bibitem[{{Fan} \& {Gibson}(2003)}]{fan03}
{Fan}, Y., \& {Gibson}, S.~E. 2003, \apjl, 589, L105

\bibitem[{{Fan} \& {Gibson}(2007)}]{fan07}
---. 2007, \apj, 668, 1232

\bibitem[{{Feynman} \& {Martin}(1995)}]{fey95}
{Feynman}, J., \& {Martin}, S.~F. 1995, \jgr, 100, 3355

\bibitem[{{Filippov}(1999)}]{fil99}
{Filippov}, B. 1999, \solphys, 185, 297

\bibitem[{{Filippov} {et~al.}(2009){Filippov}, {Golub}, \& {Koutchmy}}]{fil09}
{Filippov}, B., {Golub}, L., \& {Koutchmy}, S. 2009, \solphys, 254, 259

\bibitem[{{Fletcher} {et~al.}(2001){Fletcher}, {Metcalf}, {Alexander}, {Brown},
  \& {Ryder}}]{fle01}
{Fletcher}, L., {Metcalf}, T.~R., {Alexander}, D., {Brown}, D.~S., \& {Ryder},
  L.~A. 2001, \apj, 554, 451

\bibitem[{{Gibson} \& {Fan}(2006)}]{gib06a}
{Gibson}, S.~E., \& {Fan}, Y. 2006, Journal of Geophysical Research (Space
  Physics), 111, 12103

\bibitem[{{Golub} {et~al.}(2007){Golub}, {Deluca}, {Austin}, {Bookbinder},
  {Caldwell}, {Cheimets}, {Cirtain}, {Cosmo}, {Reid}, {Sette}, {Weber},
  {Sakao}, {Kano}, {Shibasaki}, {Hara}, {Tsuneta}, {Kumagai}, {Tamura},
  {Shimojo}, {McCracken}, {Carpenter}, {Haight}, {Siler}, {Wright}, {Tucker},
  {Rutledge}, {Barbera}, {Peres}, \& {Varisco}}]{gol07}
{Golub}, L., {Deluca}, E., {Austin}, G., {Bookbinder}, J., {Caldwell}, D.,
  {Cheimets}, P., {Cirtain}, J., {Cosmo}, M., {Reid}, P., {Sette}, A., {Weber},
  M., {Sakao}, T., {Kano}, R., {Shibasaki}, K., {Hara}, H., {Tsuneta}, S.,
  {Kumagai}, K., {Tamura}, T., {Shimojo}, M., {McCracken}, J., {Carpenter}, J.,
  {Haight}, H., {Siler}, R., {Wright}, E., {Tucker}, J., {Rutledge}, H.,
  {Barbera}, M., {Peres}, G., \& {Varisco}, S. 2007, \solphys, 243, 63

\bibitem[{{Gombosi} {et~al.}(1994){Gombosi}, {Powell}, \& {de Zeeuw}}]{gom94}
{Gombosi}, T.~I., {Powell}, K.~G., \& {de Zeeuw}, D.~L. 1994, \jgr, 99, 21525

\bibitem[{{Heinzel} {et~al.}(2008){Heinzel}, {Schmieder}, {F{\'a}rn{\'{\i}}k},
  {Schwartz}, {Labrosse}, {Kotr{\v c}}, {Anzer}, {Molodij}, {Berlicki},
  {DeLuca}, {Golub}, {Watanabe}, \& {Berger}}]{hei08}
{Heinzel}, P., {Schmieder}, B., {F{\'a}rn{\'{\i}}k}, F., {Schwartz}, P.,
  {Labrosse}, N., {Kotr{\v c}}, P., {Anzer}, U., {Molodij}, G., {Berlicki}, A.,
  {DeLuca}, E.~E., {Golub}, L., {Watanabe}, T., \& {Berger}, T. 2008, \apj,
  686, 1383

\bibitem[{{Heyvaerts} {et~al.}(1977){Heyvaerts}, {Priest}, \& {Rust}}]{hey77}
{Heyvaerts}, J., {Priest}, E.~R., \& {Rust}, D.~M. 1977, \apj, 216, 123

\bibitem[{{Isenberg} \& {Forbes}(2007)}]{ise07}
{Isenberg}, P.~A., \& {Forbes}, T.~G. 2007, \apj, 670, 1453

\bibitem[{{Karpen} {et~al.}(1998){Karpen}, {Antiochos}, {Devore}, \&
  {Golub}}]{kar98}
{Karpen}, J.~T., {Antiochos}, S.~K., {Devore}, C.~R., \& {Golub}, L. 1998,
  \apj, 495, 491

\bibitem[{{Keppens} {et~al.}(2003){Keppens}, {Nool}, {T{\'o}th}, \&
  {Goedbloed}}]{kep03}
{Keppens}, R., {Nool}, M., {T{\'o}th}, G., \& {Goedbloed}, J.~P. 2003, Computer
  Physics Communications, 153, 317

\bibitem[{{Kliem} {et~al.}(2000){Kliem}, {Karlick{\'y}}, \& {Benz}}]{kli00}
{Kliem}, B., {Karlick{\'y}}, M., \& {Benz}, A.~O. 2000, \aap, 360, 715

\bibitem[{{Kliem} \& {T{\"o}r{\"o}k}(2006)}]{kli06}
{Kliem}, B., \& {T{\"o}r{\"o}k}, T. 2006, \prl, 96, 255002

\bibitem[{{Kosugi} {et~al.}(2007){Kosugi}, {Matsuzaki}, {Sakao}, {Shimizu},
  {Sone}, {Tachikawa}, {Hashimoto}, {Minesugi}, {Ohnishi}, {Yamada}, {Tsuneta},
  {Hara}, {Ichimoto}, {Suematsu}, {Shimojo}, {Watanabe}, {Shimada}, {Davis},
  {Hill}, {Owens}, {Title}, {Culhane}, {Harra}, {Doschek}, \& {Golub}}]{kos07}
{Kosugi}, T., {Matsuzaki}, K., {Sakao}, T., {Shimizu}, T., {Sone}, Y.,
  {Tachikawa}, S., {Hashimoto}, T., {Minesugi}, K., {Ohnishi}, A., {Yamada},
  T., {Tsuneta}, S., {Hara}, H., {Ichimoto}, K., {Suematsu}, Y., {Shimojo}, M.,
  {Watanabe}, T., {Shimada}, S., {Davis}, J.~M., {Hill}, L.~D., {Owens}, J.~K.,
  {Title}, A.~M., {Culhane}, J.~L., {Harra}, L.~K., {Doschek}, G.~A., \&
  {Golub}, L. 2007, \solphys, 243, 3

\bibitem[{{Lau} \& {Finn}(1990)}]{lau90}
{Lau}, Y.-T., \& {Finn}, J.~M. 1990, \apj, 350, 672

\bibitem[{{Low} \& {Hundhausen}(1995)}]{low95}
{Low}, B.~C., \& {Hundhausen}, J.~R. 1995, \apj, 443, 818

\bibitem[{{Masson} {et~al.}(2009){Masson}, {Pariat}, {Aulanier}, \&
  {Schrijver}}]{mas09}
{Masson}, S., {Pariat}, E., {Aulanier}, G., \& {Schrijver}, C.~J. 2009, \apj,
  700, 559

\bibitem[{{Moreno-Insertis} {et~al.}(2008){Moreno-Insertis}, {Galsgaard}, \&
  {Ugarte-Urra}}]{mor08}
{Moreno-Insertis}, F., {Galsgaard}, K., \& {Ugarte-Urra}, I. 2008, \apjl, 673,
  L211

\bibitem[{{Murray} {et~al.}(2009){Murray}, {van Driel-Gesztelyi}, \&
  {Baker}}]{mur09}
{Murray}, M.~J., {van Driel-Gesztelyi}, L., \& {Baker}, D. 2009, \aap, 494, 329

\bibitem[{{Nishizuka} {et~al.}(2008){Nishizuka}, {Shimizu}, {Nakamura},
  {Otsuji}, {Okamoto}, {Katsukawa}, \& {Shibata}}]{nis08}
{Nishizuka}, N., {Shimizu}, M., {Nakamura}, T., {Otsuji}, K., {Okamoto}, T.~J.,
  {Katsukawa}, Y., \& {Shibata}, K. 2008, \apjl, 683, L83

\bibitem[{{Nistico} {et~al.}(2009){Nistico}, {Bothmer}, {Patsourakos}, \&
  {Zimbardo}}]{nis09}
{Nistico}, G., {Bothmer}, V., {Patsourakos}, S., \& {Zimbardo}, G. 2009, ArXiv
  e-prints

\bibitem[{{Pariat} {et~al.}(2009){Pariat}, {Antiochos}, \& {DeVore}}]{par09}
{Pariat}, E., {Antiochos}, S.~K., \& {DeVore}, C.~R. 2009, \apj, 691, 61

\bibitem[{{Patsourakos} {et~al.}(2008){Patsourakos}, {Pariat}, {Vourlidas},
  {Antiochos}, \& {Wuelser}}]{pat08}
{Patsourakos}, S., {Pariat}, E., {Vourlidas}, A., {Antiochos}, S.~K., \&
  {Wuelser}, J.~P. 2008, \apjl, 680, L73

\bibitem[{{Pontin} {et~al.}(2007){Pontin}, {Bhattacharjee}, \&
  {Galsgaard}}]{pon07}
{Pontin}, D.~I., {Bhattacharjee}, A., \& {Galsgaard}, K. 2007, Physics of
  Plasmas, 14, 052106

\bibitem[{{Priest} \& {Forbes}(2002)}]{pri02}
{Priest}, E.~R., \& {Forbes}, T.~G. 2002, \aapr, 10, 313

\bibitem[{{Sato} \& {Hayashi}(1979)}]{sat79}
{Sato}, T., \& {Hayashi}, T. 1979, Physics of Fluids, 22, 1189

\bibitem[{{Schmieder} {et~al.}(1995){Schmieder}, {Shibata}, {van
  Driel-Gesztelyi}, \& {Freeland}}]{sch95}
{Schmieder}, B., {Shibata}, K., {van Driel-Gesztelyi}, L., \& {Freeland}, S.
  1995, \solphys, 156, 245

\bibitem[{{Schrijver} {et~al.}(2008){Schrijver}, {Elmore}, {Kliem},
  {T{\"o}r{\"o}k}, \& {Title}}]{sch08}
{Schrijver}, C.~J., {Elmore}, C., {Kliem}, B., {T{\"o}r{\"o}k}, T., \& {Title},
  A.~M. 2008, \apj, 674, 586

\bibitem[{{Shibata} {et~al.}(1992{\natexlab{a}}){Shibata}, {Ishido}, {Acton},
  {Strong}, {Hirayama}, {Uchida}, {McAllister}, {Matsumoto}, {Tsuneta},
  {Shimizu}, {Hara}, {Sakurai}, {Ichimoto}, {Nishino}, \& {Ogawara}}]{shi92a}
{Shibata}, K., {Ishido}, Y., {Acton}, L.~W., {Strong}, K.~T., {Hirayama}, T.,
  {Uchida}, Y., {McAllister}, A.~H., {Matsumoto}, R., {Tsuneta}, S., {Shimizu},
  T., {Hara}, H., {Sakurai}, T., {Ichimoto}, K., {Nishino}, Y., \& {Ogawara},
  Y. 1992{\natexlab{a}}, \pasj, 44, L173

\bibitem[{{Shibata} {et~al.}(1994){Shibata}, {Nitta}, {Strong}, {Matsumoto},
  {Yokoyama}, {Hirayama}, {Hudson}, \& {Ogawara}}]{shi94}
{Shibata}, K., {Nitta}, N., {Strong}, K.~T., {Matsumoto}, R., {Yokoyama}, T.,
  {Hirayama}, T., {Hudson}, H., \& {Ogawara}, Y. 1994, \apjl, 431, L51

\bibitem[{{Shibata} {et~al.}(1992{\natexlab{b}}){Shibata}, {Nozawa}, \&
  {Matsumoto}}]{shi92}
{Shibata}, K., {Nozawa}, S., \& {Matsumoto}, R. 1992{\natexlab{b}}, \pasj, 44,
  265

\bibitem[{{Shibata} {et~al.}(1997){Shibata}, {Shimojo}, {Yokoyama}, \&
  {Ohyama}}]{shi97}
{Shibata}, K., {Shimojo}, M., {Yokoyama}, T., \& {Ohyama}, M. 1997, in
  Astronomical Society of the Pacific Conference Series, Vol. 111, Magnetic
  Reconnection in the Solar Atmosphere, ed. R.~D. {Bentley} \& J.~T. {Mariska},
  29--38

\bibitem[{{Titov} \& {D{\'e}moulin}(1999)}]{tit99}
{Titov}, V.~S., \& {D{\'e}moulin}, P. 1999, \aap, 351, 707

\bibitem[{{T{\"o}r{\"o}k} \& {Kliem}(2003)}]{toe03}
{T{\"o}r{\"o}k}, T., \& {Kliem}, B. 2003, \aap, 406, 1043

\bibitem[{{T{\"o}r{\"o}k} \& {Kliem}(2005)}]{toe05}
---. 2005, \apjl, 630, L97

\bibitem[{{T{\"o}r{\"o}k} \& {Kliem}(2007)}]{toe07}
---. 2007, \an, 328, 743

\bibitem[{{T{\"o}r{\"o}k} {et~al.}(2004){T{\"o}r{\"o}k}, {Kliem}, \&
  {Titov}}]{toe04}
{T{\"o}r{\"o}k}, T., {Kliem}, B., \& {Titov}, V.~S. 2004, \aap, 413, L27

\bibitem[{{Ugarte-Urra} {et~al.}(2007){Ugarte-Urra}, {Warren}, \&
  {Winebarger}}]{uga07}
{Ugarte-Urra}, I., {Warren}, H.~P., \& {Winebarger}, A.~R. 2007, \apj, 662,
  1293

\bibitem[{{van Ballegooijen}(2004)}]{van04}
{van Ballegooijen}, A.~A. 2004, \apj, 612, 519

\bibitem[{{Yokoyama} \& {Shibata}(1996)}]{yok96}
{Yokoyama}, T., \& {Shibata}, K. 1996, \pasj, 48, 353

\bibitem[{{Yokoyama} \& {Shibata}(1999)}]{yok99}
{Yokoyama}, T., \& {Shibata}, K. 1999, in American Institute of Physics
  Conference Series, Vol. 471, American Institute of Physics Conference Series,
  ed. S.~T. {Suess}, G.~A. {Gary}, \& S.~F. {Nerney}, 61--66

\end{thebibliography}

\end{document}